\documentclass[journal=nalefd,manuscript=letter]{achemso}
\usepackage{epsf}
\usepackage{graphicx}
\usepackage{dcolumn}
\usepackage{bm}
\usepackage{amsmath}
\usepackage{amsfonts}
\usepackage{amssymb}
\usepackage{ulem}

\title{Chemically-induced Mobility Gaps in Graphene Nanoribbons: A Route for
Upscaling Device Performances}

\author{Blanca Biel}
\affiliation{CEA, LETI, MINATEC, F38054 Grenoble, France}
\altaffiliation{Present address: Departamento de Electr\'onica y Tecnolog{\'\i}a de Computadores, Universidad de Granada, Campus de Fuente Nueva, and CITIC, Campus de Aynadamar, 18071 Spain}
\email{}
\author{Fran\c cois Triozon}
\affiliation{CEA, LETI, MINATEC, F38054 Grenoble, France}
\author{X. Blase}
\affiliation{Institut N\'{e}el, CNRS and Universit\'{e} Joseph Fourier, B.P. 166, 38042 Grenoble Cedex 09, France}
\author{Stephan Roche}
\affiliation{CEA, INAC/SPSMS/GT 17 rue des Martyrs, 38054 Grenoble Cedex 9, France}

\begin{document}

%
\begin{abstract}
We report an {\it ab initio}-based study of mesoscopic quantum transport
in chemically doped graphene nanoribbons with a width up to 10 nm. The occurrence
of quasibound states related to boron impurities results
in mobility gaps as large as 1 eV, driven by strong electron-hole
asymmetrical backscattering phenomena. This phenomenon opens new ways to overcome
current limitations of graphene-based devices through the fabrication of
chemically-doped graphene nanoribbons with sizes within
the reach of conventional lithography.
\end{abstract}

The basic principle of CMOS field effect transistor (CMOS-FET)
is to drive a high current density in its ON-state,
whereas electrostatic gating blocks
almost completely
the current flow in the so-called OFF-state.
The current ratio between OFF and ON states is one of the key parameters determining
the device performance \cite{CMOSFET_Sze}. This electrostatic
control of the CMOS-FET is strongly efficient owing
to the use of semiconducting materials, which present a
sufficiently large electronic bandgap to electrostatically
deplete or accumulate charges in the conducting channel.
Recently, the synthesis of low dimensional carbon
based materials has opened alternatives
to silicon-based electronics devices.
Indeed, by graphite exfoliation \cite{graphene} or
by sublimation of SiC \cite{Graphene2},
the fabrication of single layer two-dimensional (2D)
graphene or stacked graphene multilayers was achieved,
with exceptionally high values of reported charge mobilities
($\geq 100.000 \hbox{ cm}^{2}\hbox{V}^{-1}\hbox{s}^{-1}$) close to the Dirac point
\cite{graphenetransp1,graphenetransp2,graphenetransp3}.
Notwithstanding, with 2D graphene being a zero-gap semiconductor \cite{RMPG},
a poor field effect efficiency with small ON/OFF current
ratio was obtained \cite{graphenetransp1,graphenetransp2,graphenetransp3}.

E-beam lithography and oxygen plasma
etching allow to fabricate graphene nanoribbons (GNRs)
\cite{Ribbonpure1,Ribbonpure2,Ribbonpure4},
down to a few tens of
nanometers in width, 
opening the possibility for bandgap engineering
through electronic confinement.
However, current limitations of state-of-the-art
lithographic techniques hardly allow to fabricate GNRs with sufficiently
large energy bandgaps \cite{KimENRJGAP1,KimENRJGAP2,IBM1,IBM2,IBM3}.
To further reduce the GNRs lateral size down to a
few nanometers, alternative chemical approaches
have been developed and competitive device performances
have been reported \cite{Dai1,Dai2}. 
These approaches remain however hardly
compatible with massive integration of interconnected
devices at the wafer scale.

Chemical doping allows
to develop new applications such as chemical sensors or more efficient 
electrochemical switches \cite{AMO1,AMO2}. 
The conductivity modulation due to chemical
doping and functionalization of 2D graphene has been
explored numerically \cite{DF1,DF2},
without showing significant improvement
of device performances.
Several theoretical studies have also investigated the effect
of doping on the properties of single-layer GNRs.
The impact of a single boron impurity on the electronic
properties of GNRs was analyzed by first-principles methods
in Refs. \cite{Fazzio-Bdoped,BBTR}. 
Other works have considered
the GNRs bandgap tuning by doping or functionalization \cite{dopedGNRs1,dopedGNRs2,dopedGNRs3}, but
only gaps of a few tens of meV have been predicted.
Theoretically, two
types of GNRs with highly symmetric edges have been
described, namely, zigzag (zGNRs) and armchair (aGNRs).
To date, most efforts have focused
on the spin-transport properties of zGNRs,
while the possibilities for
charge transport of doped aGNRs have somehow been neglected. 
Besides, due to the computational cost of \textsl{ab initio} calculations,
the width of the GNRs studied in those works is in general well below 3 nm.

In this Article, we report on a quantum transport
study of chemically doped GNRs up to 10 nm in width.
By combining first-principles calculations with tight-binding models, the effect of chemical doping
on charge conduction is explored
for ribbons with a length up to the micron
and random distribution of
substitutional boron impurities. The joint effects of backscattering
together with the enhanced localization due to low dimensionality
are shown to produce
strong electron-hole transport asymmetry and large
mobility gaps, even in the low doping limit. These chemically modified
graphene-based materials, within reach of current
lithographic techniques, should therefore
allow for a strongly enhanced current
density modulation under electrostatic gating.

We focus on the case of aGNRs.
This type of ribbon symmetry has theoretically
been proposed as energetically more stable
and more resistant to edge defects when
compared to zGNRs \cite{Kawai,Hakkinen,Wassmann}, which seems to be confirmed by
experimental observations \cite{armchair_exp}. As reported by \textsl{ab initio}
calculations \cite{Louie-gaps, Barone-gaps, White-gaps},
aGNRs are found to be always semiconducting
with an energy gap that varies up to 0.5 eV for ribbons of similar width,
depending on the exact number of dimer chains within the unit cell.
Unlike zGNRs, the ground
state of aGNRs is not spin-polarized.
{\bf\it p}-type ({\bf\it n}-type)
doping can be achieved
by the incorporation of boron (nitrogen)
atoms in substitution within the $sp^2$-type carbon matrix, 
as reported in
Ref. \cite{BdopedCVD1} for carbon nanotubes and very recently
in Ref. \cite{Rao} in the case of 2D graphene.
First-principles
calculations of the effect of a single boron 
impurity on the electronic properties of aGNRs have been
reported in Ref. \cite{BBTR} 
for aGNRs with widths between 2.3 and 4.2 nm.
Following Ref. \cite{Louie-gaps}, we refer to a aGNR with
N dimers contained in its unit cell as an N-aGNR.
Both types of possible electronic structures,
namely the pseudo-metallic ribbons 20-aGNR and 35-aGNR (i.e., ribbons predicted
to be metallic within a simple nearest-neighbor tight-binding model), 
and the semiconducting 34-aGNR were addressed.
The modification of the electronic structure of the aGNRs by
a single boron (or nitrogen) impurity
turns out to be strongly dependent on
the position of the dopant with respect to the ribbon edge,
displaying a significant electron-hole asymmetry
\cite{Louie_asym,AvourisNL}.
This phenomenon results from the interplay between wavefunction
symmetries and screening effects that are system-dependent.
Regardless of their exact position across the ribbon width,
boron impurities are found to preserve their well-known acceptor
character (or hole-doping ability) in carbon, with vanishingly weak
impact of the transport properties in the first conduction
band close to the Fermi level (E$_F$), which almost preserves conductance quantization. 

The large
variation of resonant energies with the acceptor (donor) chemical impurity
position allows however to infer
that a random distribution of impurities
will likely result in an
uniform reduction of conductance over the occupied (unoccupied)
states part of the first conduction {\it plateau},
yielding marked electron-hole asymmetry and charge mobility
gaps intrinsic to p-type (n-type) doping.
The present work is the first theoretical study based on {\it ab initio} atomistic models
providing confirmation to such prediction.

Due to the large size of our unit cells, a statistical study of 
the mesoscopic transport using the fully {\it ab initio} Hamiltonian \cite{Gomez-Navarro}
is not possible here. We have used instead a combination of first-principles
calculations with tight-binding models, that has been satisfactorily employed before
for doped carbon nanotubes \cite{Avriller}.
As a first step, the scattering potential induced by the dopant
has been extracted from those first-principles density functional theory (DFT)
calculations.
The analysis of the
\textsl{ab initio} onsite and hopping terms around the dopant, as obtained from
self-consistent DFT calculations within
an atomic-like basis,
allows to build a simple nearest-neighbor tight-binding (nn-TB) model.
The substitutional impurity modifies mainly the onsite terms, creating
a potential well on a typical length scale of about 10 Angstrom (see
Ref. \cite{Adessi} and Ref. \cite{DF1} for a detailed analysis).
\begin{figure}[htp]
\centering
\includegraphics[width=10cm]{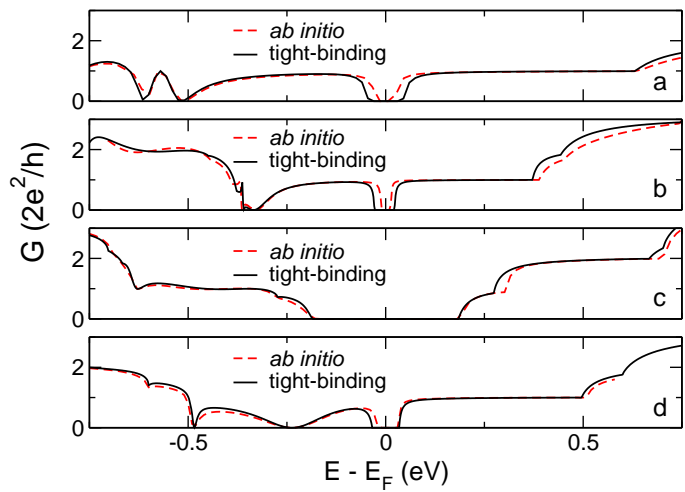}
\caption{Conductance profile as a function of
energy for the single-dopant case, both
within the full \textsl{ab initio}
approach (dashed red line) and the tight-binding model (solid black line)
for the (a) 20-, (b) 35-, (c) 31- and (d) 26-aGNRs;
the dopant has been placed at the (a) center,
(b) midway between center and edge, (c) first neighbor to atom at
edge, and (d) edge positions, respectively.}
\label{FIG1}
\end{figure}
A modification of the C-C TB hopping at the edges is also
included in the TB Hamiltonian following Ref. \cite{Louie-gaps} to simulate
the edge passivation, that yields a small energy gap
at the charge neutrality point (CNP). As shown in \ref{FIG1}, the conductance
for a single dopant (in various ribbon symmetries and
impurity location) computed by our TB models nicely
reproduces the \textsl{ab initio} transport calculations performed with
the TABLIER code \cite{Adessi} for an extensive set of dopant positions 
and ribbons width.
The conductance of long GNRs is then calculated within the
Landauer-B\"uttiker formalism
\cite{Datta}.
Order(N) decimation techniques are
used to explore the various transport properties
for randomly doped ribbons as long as 1\ $\mu$m.
By combining the accuracy of first-principles to
describe the local electronic structure with
the numerical efficiency of a nn-TB model,
our transport methodology thus allows for a realistic
exploration of the mesoscopic transport in
randomly doped GNRs.

We first analyze the impact of different doping rates on ribbons of about
4 nm in width, namely the pseudo-metallic 35-aGNR and the semiconducting
34-aGNR. Ribbons with a length up to 1 $\mu$m are considered,
and impurities are uniformly distributed over the whole ribbon length and width,
with a restriction preventing the overlap of the scattering potentials of individual
dopants.
\begin{figure}[htp]
\centering
\includegraphics[width=9cm]{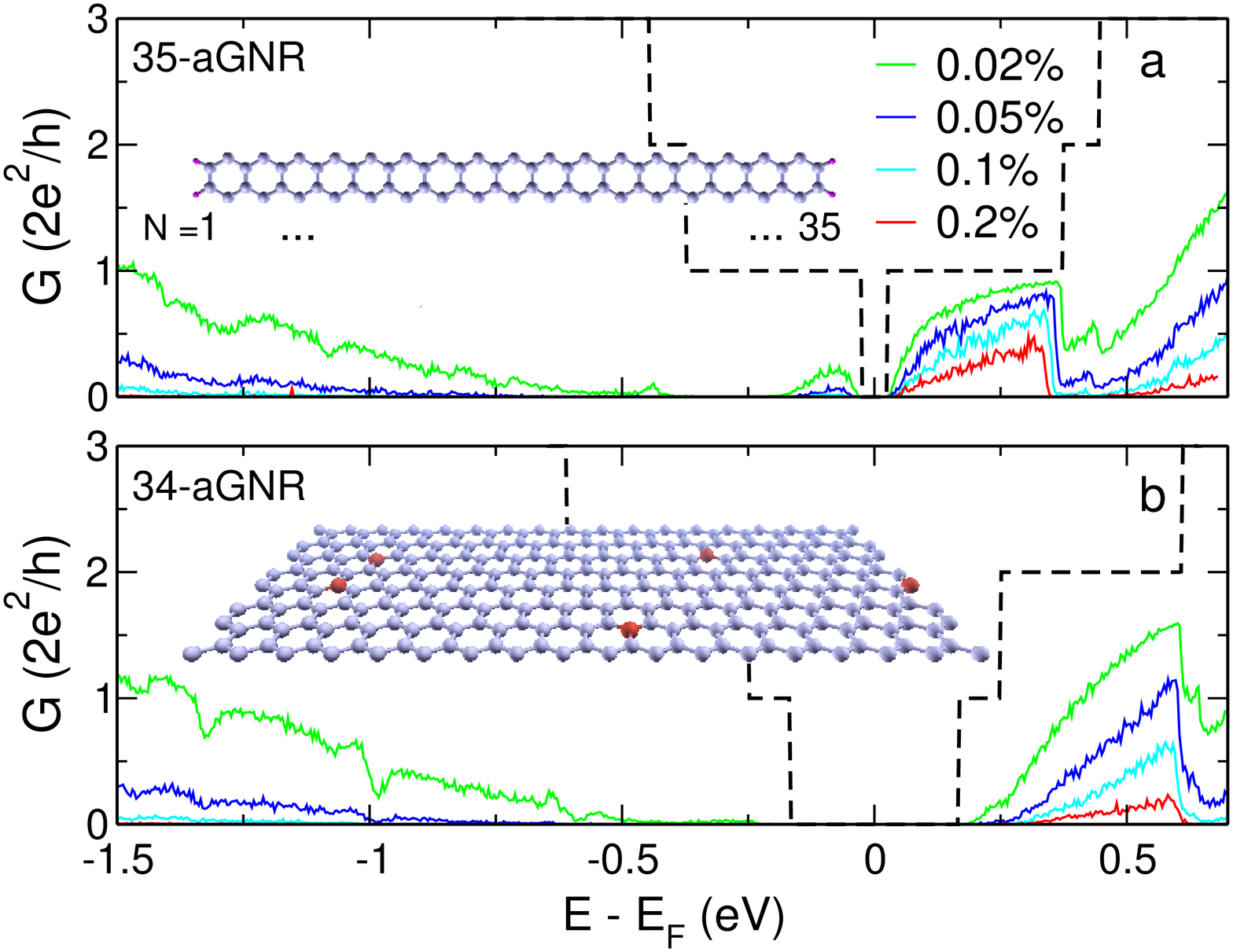}
\caption{(a) Average conductance as a function of energy for the pseudo-metallic 35-aGNR
for doping rates $\approx$
0.02\%, 0.05\%, 0.1\% and 0.2\% (from top to bottom). The dashed black line
corresponds to
the ideal (undoped) case.
The averages have been performed over $\approx$ 500 disorder
realizations with a ribbon length of $\approx$ 1 $\mu$m.
The inset shows the unit cell of the 35-aGNR with passivating H atoms.
N indicates the number of dimer chains in a N-aGNR.
(b) Same as in (a, main frame) for the semiconducting 34-aGNR.
Inset: Schematic view of a randomly doped GNR.}
\label{FIG2}
\end{figure}
\ref{FIG2} shows the conductance as a function of energy for
the (a) 35- and (b) 34-aGNRs, for doping rates between $\approx$ 0.02\% and 0.2\%.
In this formalism, electrodes are treated as semi-infinite, perfect GNRs, and
disorder is included only in the region between
the electrodes (channel). Since the scattering potential induces an overall shift of
the CNP within the channel,
the CNP of electrodes and channel are not aligned, leading to a displacement of the
doping-induced
gap with respect to that of the perfect ribbon. To properly analyze our results we
have thus simply
proceeded to a rigid shift of the conductance profile of the doped system to
superimpose it to that of the ideal case.
As a result of the acceptor-like character
of the impurity states induced by the boron dopant,
the conductance is affected in a clear asymmetric
fashion for energy values below or 
above the CNP \cite{BBTR}, as evidenced by the opening
of a large mobility gap that extends well beyond the
first conductance \textsl{plateau} below
the small initial electronic bandgap.
The mobility gap width reaches almost 1 eV,
of the order of the silicon energy gap.

Interestingly, despite
the smaller size of the initial gap of the
pseudo-metallic 35-aGNR with respect to that of the 34-aGNR, 
the magnitude of the doping-induced mobility gap in both ribbons
is very similar for the same ribbon length and doping rate.
This fact, as well as the higher conductance values for electron transport
in the 35-aGNR (due to the robustness against
scattering processes of the linear antibonding $\pi^*$ band 
in pseudo-metallic ribbons), make them even more suitable for their use in
electronic devices.
\begin{figure}[htp]
\centering
\includegraphics[width=9cm]{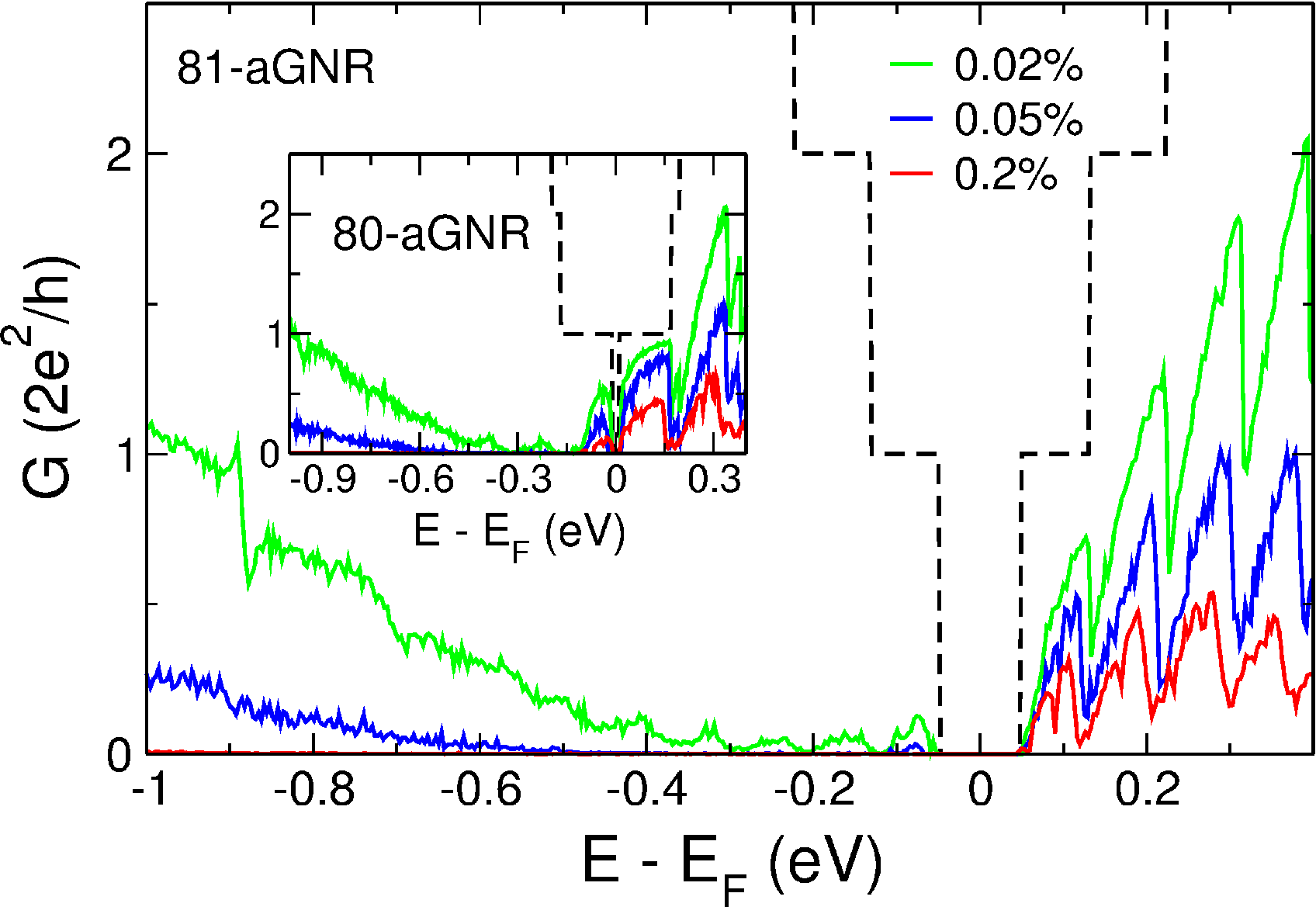}
\caption{Main panel: Same as \ref{FIG2} for the semiconducting 81-aGNR
and three selected doping rates ($\approx$ 0.02\%, 0.05\% and 0.2\%, from top
to bottom). Inset: Same as in main frame for the pseudo-metallic 80-aGNR.}
\label{FIG3}
\end{figure}

Since the width of these GNRs is reaching the capability limits of
standard lithographic techniques, we further explore the case of aGNRs with a width
of $\approx$ 10 nm, namely the 80- and the 81-aGNR. For these systems, a doping
rate below 0.2\% is enough to achieve a mobility gap of $\approx$ 1 eV 
(see \ref{FIG3}, while keeping high conductance values in the conduction
band for both types of wide GNRs.
In this case, higher doping rates are required to achieve
the full suppression of conductance
in the valence band. This is due to the fact that
backscattering in aGNRs is highly dependent on the dopant position,
with a maximum impact when the dopant is close to (or right at) the
ribbon edge \cite{BBTR}. As a consequence of
a uniform random distribution of dopants,
the probability to find an impurity close to the edges
decreases with increasing ribbon width, and will be,
for a 80-aGNR, almost one half than for a 35-aGNR, 
thus leading to a lower impact on the conductance for a similar doping rate.
\begin{figure}[htp]
\centering
\includegraphics[width=9cm]{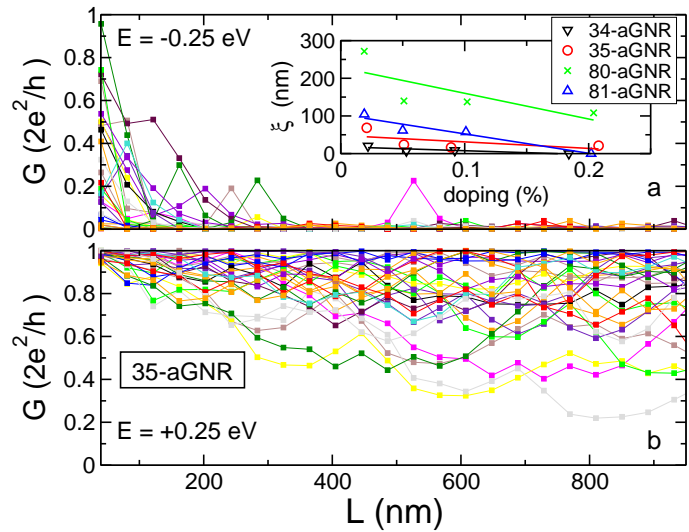}
\caption{Main frame: Conductance as a function of length for a
set of doping realizations of the 35-aGNR at 0.25 eV below (a)
and above (b) the CNP for a doping rate of $\approx$ 0.05\%. Curves
with same color in (a) and (b)
correspond to the same position of the dopants along the ribbon.
Inset: Estimated localization lengths as a function of doping rate for the 34- and
35-GNRs (at E = -0.2 eV), and the 80- and 81-aGNRs
(at E = -0.1 eV).  Solid lines are fittings to the calculated values.}
\label{FIG4}
\end{figure}

The asymmetry in the electron {\it vs}.
hole conduction \cite{Louie_asym,AvourisNL} is also spectacularly 
evidenced by scrutinizing at the energy-dependent
transport regimes. \ref{FIG4} (main frame) shows 
the length dependence of the conductance for selected individual 
randomly doped 35-aGNRs, with a doping rate of $\approx$ 0.05\%,
at a energy = 0.25 eV below (a)
and above (b) the CNP.
For energy values lying in the conduction band (b), 
the conductance for various random configurations is
seen to slowly decay with ribbon length with values staying
close to the conductance quantum ($G_{0}=2e^{2}/h$),
pinpointing a quasi-ballistic regime.
Though all 1-dimensional systems show localization in the presence of disorder,
for lower doping rates quantum interference effects are too weak to allow localization
to manifest itself in the length scale considered in this study.
In contrast,
for energies lying in the valence band,
a strong exponential decrease
of the conductance is found (a),
with large fluctuations associated
to different defects positions. The exponential
decay of the conductance with
the length of the ribbon is related to the Anderson
localization, that has already been observed experimentally
at room temperature in defected metallic carbon nanotubes
\cite{Gomez-Navarro}, due to their long phase coherence lengths.

In the inset of \ref{FIG4},
the estimated localization lengths ($\xi$) for
hole transport at different doping rates are reported.
A statistical average over
more than 500 disorder samples is performed
to extract the $\xi$ from $\langle\ln G/G_{0}\rangle=-\xi/L$,
with $L$ the ribbon length. 
The localization length ranges
within 10-300 nm, depending on the ribbon width and doping density,
and is inversely proportional to the doping density.
$\xi$ is also found to further increase with the ribbon width but in a
non-linear fashion. This comes from the non-uniformity
of the disorder potential with the dopant position.

In the case of carbon nanotubes,
the localization length is predicted to increase
linearly with the size of the system \cite{White_Nature},
a trend confirmed experimentally \cite{Flores}.
A similar scaling law was derived in pseudo-metallic
GNRs \cite{White_transport}. This can be explained by invoking
a generalization of Thouless relation for quasi-one dimensional
systems, $\xi \sim N_{\perp} l_e$,
where $N_{\perp}$ is the number of conducting modes available at a certain energy and
$l_e$ is the elastic mean free path.
In the first \textsl{plateau} around the CNP,
ribbons of different widths
present the same number of channels ($N_{\perp}$ = 1),
and $l_e$ is predicted to increase with
the number of atoms in the unit cell \cite{White_transport},
thus leading to an increase of $\xi$. However, this is true only
in the presence of a uniform disorder potential, which is not the case here. We note
that similar power law scaling of $\xi$ has
been observed for the case of edge disorder \cite{Heinzel,Mucciolo}.

In conclusion, we have performed a numerical
transport study of boron doped armchair GNRs with widths up to $\approx$ 10 nm,
and lengths reaching the micron scale and doping rates from 0.02\% to 0.2\%.
Depending on the energy of charge carriers, the transport can vary from quasiballistic
to a strongly localized regime, as a result of a strong electron/hole asymmetry
induced by
the chemical doping. This phenomenon leads
to mobility gaps in the order of 1 eV,
while the conductance in the conduction bands remains large,
with values close to those of the ideal ribbon for low enough doping rates.
This opens an unprecedented way to improve the
performances of graphene-based devices since the obtained
asymmetry should also manifest in a considerably higher ON/OFF current ratio.
One notes that electron-hole conduction asymmetry 
in graphene devices by chemical functionalization has been recently reported by
Farmer and coworkers \cite{AvourisNL},
whereas Raman studies of graphene-FETs allow
for efficient monitoring of doping effects \cite{Das}.

\acknowledgement

This work was supported by the GRAPHENE project of CARNOT Institute (LETI) and
the ICT/FET European funding from GRAND and ANR-06-NANO-069-02 "Accent" projects.
We thank the CEA/CCRT supercomputing facilities for providing computational resources. 

\bibliography{achemso}

\end{document}